\documentclass[a4paper]{jpconf}
\usepackage{graphicx}

\newcommand{\cM}{\mathcal{M}}

\begin{document}
\title{Reference calculations for subthreshold $\Xi$ production}

\author{Boris Tom\'a\v{s}ik$^{a, b}$ and Evgeni E Kolomeitsev$^a$}

\address{$^a$ Univerzita Mateja Bela, FPV, Tajovsk\'eho 40,
97401 Bansk\'a Bystrica, Slovakia}
\address{$^b$ Czech Technical University in Prague, FNSPE, B\v{r}ehov\'a 7,
11519 Prague 1, Czech Republic}

\ead{boris.tomasik@cern.ch}

\begin{abstract}
We present a minimal statistical model designed for the description
of rare-hadron multiplicities in nucleus-nucleus collisions at energies below the threshold of  the particle production in binary elementary collisions.
Differences to more conventional canonical statistical model are explained.
The minimal statistical model is applied to the description of
multiplicity ratios involving $\Xi$ hyperons, which are measured by the HADES
collaboration at GSI-SIS. It is argued that the HADES data cannot be reproduced by the model based on the statistical equilibrium and the strangeness conservation. The data remain underpredicted even when in-medium potentials acting on hadrons are taken into account.
This hints to non-equilibrium production of the $\Xi$ hyperons and their continuous freeze-out.
\end{abstract}


\section{Introduction}

Data obtained by HADES collaboration~\cite{hades_xi} show that doubly strange $\Xi$ hyperon can be produced in nuclear collisions at energies well below the production threshold for nucleon-nucleon collisions. In Ar+KCl collisions at $E_{\mathrm{beam}}= 1.76$~$A$GeV the
observed ratio of $\Xi$ to $(\Lambda + \Sigma^0)$ multiplicities is
$\cM_\Xi/\cM_{\Lambda+\Sigma^0} = (5.6 \pm 1.2^{+1.8}_{-1.7})\times
10^{-3}$, and some $\Xi$'s are also seen in Au+Au collisions at 1.23~$A$GeV.
Statistical model with canonical suppression of strangeness production
predicts a value of the order $(10^{-4})$ for the ratio.


\section{Isospin symmetry and total strangeness content}

The colliding nuclei with the atomic number $A$ have usually different number of protons ($Z$) and neutrons ($A-Z$) and thereby the matter created in their collisions has a non-zero total isospin projections quantified by the asymmetry coefficient $\eta = (A-Z)/Z$ being different from one. In our calculations we assume that this asymmetry is reflected in the observed multiplicities of hadrons belonging to one isospin multiplet, e.g,
\begin{equation}
\frac{\cM_{K^0}}{\cM_{K^+}} = \frac{\cM_{K^-}}{\cM_{\bar K^0}} =
\frac{\cM_{\Xi^-}}{\cM_{\Xi^0}}
= \frac{\cM_{n}}{\cM_{p}} = \eta.
\label{e:iso}
\end{equation}
For the Ar+KCl collisions we have $\eta \approx 1.14$.

We can use the relations (\ref{e:iso}) to determine the total amount of produced $s\bar s$
quark pairs (distributed among hadrons). In a baryon-rich environment, created in collisions at SIS energies, the $\bar s$ antiquarks can only be accommodated in kaons. Furthermore,
under such conditions kaons do not find partners for inelastic scattering
and once produced they are guaranteed to leave the system. Thus the
number of $s\bar s$ pairs can be reconstructed as
\begin{equation}
{\cal M}_{s\bar s} = {\cal M}_{K^+} + {\cal M}_{K^0} = (1 + \eta) {\cal M}_{K^+}\,  .
\end{equation}
Once the numbers of produced $s$ and $\bar{s}$ quarks are known we assume in our minimal statistical model~\cite{ourpaper} that the $s$ quarks are distributed among strange hadrons (anti-kaons and hyperons) with relative abundances  determined by thermal and chemical equilibria and by the strangeness conservation. For our analysis we use the ratios which formally do not depend on the produced strangeness~\cite{ourpaper}
\begin{eqnarray}
&&\frac{\mathcal{M}_{K^-}}{\mathcal{M}_{K^+}}=2.54^{+1.21}_{-0.91}\times
10^{-2}\,, \quad
\frac{\mathcal{M}_{\Lambda+\Sigma^0}}{\mathcal{M}_{K^+}}=1.46^{+0.49}_{-0.37}
\nonumber\\
&&\frac{\mathcal{M}_{\Sigma^++\Sigma^-}}{2\,\mathcal{M}_{K^+}}=0.30^{+0.23}_{-0.17}
\,,\quad
\frac{{\cal M}_{\Xi^-}}{{\cal M}_{K^+}{\cal M}_{\Lambda+\Sigma^0}} =
0.20^{+0.16}_{-0.12}\,.
\label{e:data}
\end{eqnarray}
These ratios has the same total number of strange an anti-strange quarks in both numerator and denominator.


\section{The minimal statistical model}

At subthreshold energies no strange particles are created in incident
nucleon-nucleon collisions and all strangeness must be produced
\emph{during} the evolution of the hot fireball created in the collision. Creation of strangeness is a very rare event, with $K^+$ multiplicity
\begin{equation}
{\cal M}_{K^+} = (2.8 \pm 0.4) \times 10^{-2} \, .
\end{equation}
The amount of $s\bar s$ pairs is thus far below the
saturation and can be treated perturbatively. It
will not only scale with the volume, but also we can assume that it will steadily grow
with time. If --- for simplicity --- we assume ideal hydrodynamics,
then the only length scale is given by $V^{1/3}$, where $V$ is the fireball volume, and it must determine
also the lifetime~\cite{RI92}. At fixed impact parameter, the probability of
creating one $s\bar s$ pair is given as
\begin{equation}
\label{e:WV}
W = \lambda V^{4/3}\,  .
\end{equation}
For a fixed volume $V$, the probability to have $n$ of
the $s\bar s$ pairs is Poissonian
\begin{equation}
P_{s\bar s}^{(n)} = e^{-W}\frac{W^n}{n!}\,  .
\end{equation}

The HADES data have been obtained after averaging over collisions with different centralities.
The dependence on $V^{4/3}$ in~(\ref{e:WV}) makes the averaging
non-trivial. If we denote the centrality averaging by angle brackets,  $\langle\cdots\rangle$,
we can relate the parameter $\lambda$ to the measured kaon multiplicity as follows
\begin{equation}
\lambda = \frac{\langle W\rangle}{\langle V^{4/3}\rangle}
= \frac{{\cal M}_{s\bar s}}{\langle V^{4/3}\rangle}
= \frac{{\cal M}_{K^+} (1 + \eta)}{\langle V^{4/3}\rangle }\, .
\end{equation}

In the minimal statistical model we assume that in an event with volume
$V$ and $n$ pairs of $s\bar s$ quarks, the $s$ quarks are distributed into
hadrons according to the probability
\begin{equation}
P_a^{(n)} = \big( z_s^{(n)}\big)^{s_a} V\, p_a =
\big( z_s^{(n)} \big)^{s_a} V\,e^{B_a\mu_B/T} \frac{m^2_a T}{2\pi}
K_2\left (\frac{m_a}{T}\right )\, .
\label{e:Pa}
\end{equation}
This is the probability to find hadron $a$ with the mass $m_a$, baryon number $B_a$ and
strangeness content $s_a$. The chemical potential is determined by the fireball freeze-out density taken $0.6\rho_0$, where $\rho_0=0.16$\,fm$^{-3}$ is the nuclear saturation density. The probability (\ref{e:Pa}) is normalised --- for
each $n$ separately --- by the factor $z_s^{(n)}$.

In total, average multiplicity of hadron species $a$ is then given by the sum
\begin{equation}
\cM_a = \langle M_a^{(1)}\rangle + \langle M_a^{(2)} \rangle + \langle M_a^{(3)} \rangle
+ \dots
\end{equation}
where the r.h.s.\ lists all contributions to the average multiplicity from events with one,
two, three, etc.\ $s\bar s$ pairs. Note that $M_a^{(n)}$ is understood as the number of hadrons $a$
produced in events with $n$ $s\bar s$ pairs, divided by the number of all events. These
multiplicities are determined with the help of $P_a^{(n)}$ and $P_{s\bar s}^{(n)}$. The
$\Xi$ hyperons, caring two $s$ quarks, are produced only in events with two and more $s\bar s$ pairs.


\section{The results}

\begin{figure}[t]
\centerline{\includegraphics[width=0.95\textwidth]{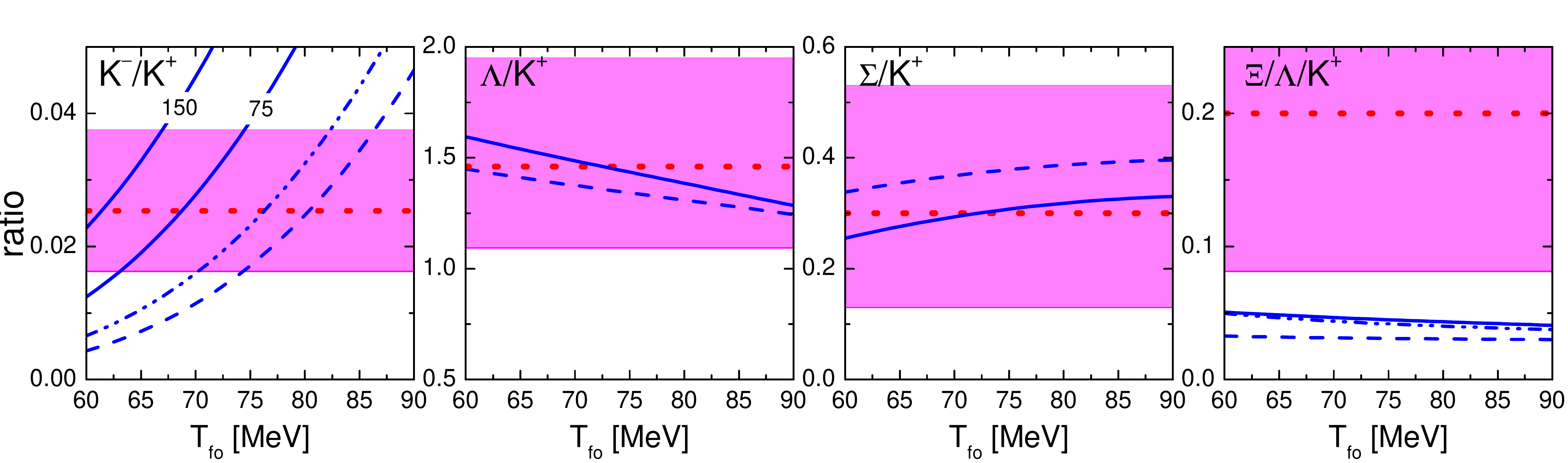}}
\caption{\label{f:rats}
Multiplicity ratios as functions of the temperature. Red dotted lines with purple bands show
the measured data values~(\ref{e:data}). Dashed lines are minimal
statistical model predictions according to~(\ref{e:ratio}). Solid lines are predictions
with in-medium potentials included. For the $\cM_{K^-}/\cM_{K^+}$ ratio we show
results with two values of $\bar K$ optical potential acting on $\bar U$: $-75$ and
$-150$~MeV (other curves are calculated with $-75$~MeV). Dash-doubly-dotted
curves show ratios with only nucleon potentials included.
}
\end{figure}

Our result for the $\Xi$ multiplicity ratio is
\begin{equation}
\frac{\cM_\Xi}{\cM_{\Lambda + \Sigma^0}  \cM_{K^+}} =
\frac{\big\langle V^{5/3} \big\rangle\langle V \rangle}%
{2\big\langle V^{4/3} \big\rangle^2}
\eta \frac{ p_{\Xi}/(p_{\bar K} + p_\Lambda + p_\Sigma)} {\langle V \rangle
\big(p_\Lambda+\frac{\eta p_\Sigma}{\eta^2+\eta+1}\big)}\,  .
\label{e:ratio}
\end{equation}
In the pre-factor we collected the modifications with respect to the ``usual'' statistical
model with the canonical suppression.
This and all other calculated ratios are displayed in Fig.~\ref{f:rats} by dashed lines
as functions of temperature. We see that the calculated
${\cM_\Xi}/{\cM_{\Lambda + \Sigma^0}  \cM_{K^+}}$ is clearly below the measured
error band.
In Fig.~\ref{f:rats} we also revised the data on $\Sigma/K^+$ ratio based on the actual
isospin ratio from~(\ref{e:iso}), see~\cite{ourpaper}.

By trying to improve the agreement of our calculations with the data we have introduced
density-dependent scalar and vector potentials $S_a$ and $V_a$~\cite{ourpaper}. They modify the statistical distribution function by replacing the dispersion relation to $E_a(p) = \sqrt{(m_a+S_a)^2 + p^2} + V_a$.
The chosen values were
$V_N = V_\Delta = \frac{3}{2}V_\Lambda=\frac{3}{2}V_\Sigma = 3 V_\Xi  =
130\,\mbox{MeV}{\rho_B}/{\rho_0}$, and
$S_N  =  -190\, \mbox{MeV}{\rho_B}/{\rho_0}$.
The scalar potentials for strange particles follow from the relation
$S_a = (U_a - V_a(\rho_0)){\rho_B}/{\rho_0}$, where we use
$U_\Lambda = -27\,\mbox{MeV}$,
$U_\Sigma = 24\,\mbox{MeV}$,
$U_\Xi = -14\,\mbox{MeV}$,
$U_{\bar K} = -75$ or $-150\, \mbox{MeV}$ and $V_{\bar{K}}=0$.
This leads to changes in our theoretical results which allow to explain $K^-/K^+$, $\Lambda/K^+$ and $\Sigma/K^+$ ratios by one freeze-out temperature $\sim 70$\,MeV. However, the changes are not large enough to bring the $\Xi/\Lambda/K^+$ ratio into agreement with the measured data.

It is interesting that accounting for the trigger on central  collisions actually makes the
disagreement even bigger. The culprit is in the volume factors which \emph{reduce}
the overall result when selection on larger volumes is made.


\section{A prediction for Au+Au collisions at 1.23~GeV}

Although the model disagrees with the Ar+KCl data, for curiosity we can make a
prediction also for the Au+Au collisions recently recorded by HADES. As seen from
Fig.~\ref{f:rats}, the temperature dependence of
$\cM_\Xi/\cM_{\Lambda + \Sigma^0}  \cM_{K^+}$ is very weak. The crucial difference
to the smaller system is in the volume which is almost five times bigger for Au+Au than for
Ar+KCl. As the ratio (\ref{e:ratio}) is proportional to $V^{-1}$, we expect that it will be
five-times smaller in gold-induced reactions than in Ar+KCl. Below the strangeness production
threshold this statement is independent of the collision energy.


\section{Conclusions}

In comparison to the canonical statistical model, the minimal statistical model
assumes that strangeness must be produced during the fireball evolution
and takes into account the proper scaling of rare species multiplicity with the volume. This
leads to non-trivial differences between the two models and the result of the
minimal statistical model on centrality-averaged multiplicities depends on a set of
averaged volume factors.

In any case, the conclusion we could draw from our study is that cascade production is non-statistical. We speculate that it might be caused by immediate decoupling of the produced $\Xi$'s from the bulk
matter. At an early hot stage some $\Xi$'s are produced which are later not annihilated
at lower temperature. Those particles chemically decouple from the system.

We speculate that the most viable for such a production of $\Xi$ would be the
strangeness recombination reactions, either involving $\bar K$ and a hyperon, or two
hyperons. Note that we have provided parametrised cross-sections for such reactions
in \cite{SQMproc}.
An interesting options appears in the $\bar K \Lambda$ channel, which in vacuum
has a high reaction threshold of 1609~MeV. Due to effective lowering of kaon
mass in the baryon-rich environment it can decrease so that eventually it hits
the $\Xi^*$ resonance around 1530~MeV. This could increase the cross section for
$\Xi$ production considerably.

Kinetic calculations aiming the the validation of these mechanisms are planned
for the near future.


\subsection*{Acknowledgements}

This work was partially supported by APVV-0050-11, VEGA 1/0469/15 (Slovakia) and
M\v{S}MT grant LG13031 (Czech Republic). We also want to acknowledge the support
towards attendance of this conference 
by the Plenipotentiary of the Slovak Government to JINR Dubna.


\end{document}